\documentclass[11pt,preprint]{aastex}
\def\gtsima{$\; \buildrel > \over \sim \;$}
\def\gsim{\lower.5ex\hbox{\gtsima}}

\newdimen\minuswidth    
\setbox0=\hbox{$-$}
\minuswidth=\wd0
\catcode`@=\active
\def@{\kern\minuswidth}
\setbox0=\hbox{\rm0}
 
\shorttitle{} 
\shortauthors{Lanzoni et al.}
 
\begin{document} 
\title{The ESO Multi-Instrument Kinematic Survey (MIKiS) of Galactic
  Globular Clusters: solid body rotation and anomalous velocity
  dispersion profile in NGC 5986\footnote{Based on FLAMES and KMOS
    observations performed at the European Southern Observatory as
    part of the Large Programme 193.D-0232 (PI: Ferraro).}}

 \author{B. Lanzoni\altaffilmark{2,3},
   F. R. Ferraro\altaffilmark{2,3}, A. Mucciarelli\altaffilmark{2,3},
   C. Pallanca\altaffilmark{2,3}, M. A. Tiongco\altaffilmark{4},
   A. Varri\altaffilmark{5}, E. Vesperini\altaffilmark{4},
   M. Bellazzini\altaffilmark{3}, E. Dalessandro\altaffilmark{3},
   L. Origlia\altaffilmark{3}, E. Valenti\altaffilmark{6},
   A. Sollima\altaffilmark{3}, E. Lapenna\altaffilmark{2,3},
   G. Beccari\altaffilmark{6} }
\affil{\altaffilmark{2} Dipartimento di Fisica e Astronomia,
  Universit\`a degli Studi di Bologna, via Gobetti 93/2, I$-$40129
  Bologna, Italy}
\affil{\altaffilmark{3}INAF-Osservatorio di Astrofisica e Scienza
  dello Spazio di Bologna, Via Gobetti 93/3, 40129, Bologna, Italy}
\affil{\altaffilmark{4} Department of Astronomy, Indiana University,
  Bloomington, IN, 47401, USA }
\affil{\altaffilmark{6} Institute for Astronomy, University of
  Edinburgh, Royal Observatory, Blackford Hill, Edinburgh EH9 3HJ, UK
\affil{\altaffilmark{6}European Southern Observatory,
  Karl-Schwarzschild-Strasse 2, 85748 Garching bei M\"{u}nchen,
  Germany}
}

\date{3 August 2018, ApJ in press}

\begin{abstract}
As part of the ESO-VLT Multi-Instrument Kinematic Survey (MIKiS) of
Galactic globular clusters, we present a detailed investigation of the
internal kinematics of NGC 5986.  The analysis is based on about 300
individual radial velocities of stars located at various distances
from the cluster center, up to $300\arcsec$ (about 4 half-mass radii).
Our analysis reveals the presence of a solid-body rotation extending
from the cluster center to the outermost regions probed by the data,
and a velocity dispersion profile initially declining with the
distance from the cluster's center, but flattening and staying
constant at $\sim 5$ km s$^{-1}$ for distances larger than about one
half-mass radius.  This is the first globular cluster for which
evidence of the joint presence of solid-body rotation and flattening
in the outer velocity dispersion profile is found. The combination of
these two kinematical features provides a unique opportunity to shed
light on fundamental aspects of globular cluster dynamics and probe
the extent to which internal relaxation, star escape, angular momentum
transport and loss, and the interaction with the Galaxy tidal field
can affect a cluster's dynamical evolution and determine its current
kinematical properties. We present the results of a series of N-body
simulations illustrating the possible dynamical paths leading to
kinematic features like those observed in this cluster and the
fundamental dynamical processes that underpin them.
\end{abstract}
 

\section{INTRODUCTION}
\label{sec_intro}
The ESO-VLT Multi-Instrument Kinematic Survey (hereafter the MIKiS
survey; \citealp{ferraro+18a}) of Galactic globular clusters (GCs) is
a project specifically designed to characterize the kinematical
properties along the line-of-sight of an illustrative selection of GCs
in the Milky Way. We have measured individual radial velocities (RVs)
of hundreds of stars, distributed over the entire radial range of each
stellar system. To this end, the survey fully exploits the
spectroscopic capabilities of different instruments currently
available at the ESO Very Large Telescope (VLT): the adaptive-optics
assisted integral-field spectrograph SINFONI, the multi-object
integral-field spectrograph KMOS, and the multi-object fiber-fed
spectrograph FLAMES.

The evolutionary interplay between two-body relaxation-driven
processes and the effects of the external tidal environment gives
origin to a rich internal kinematics in collisional systems, which is
now observationally accessible.  Moreover, Milky Way GCs are very old
stellar systems (with ages of $\sim 10$ Gyr or more; see
e.g. \citealt{forbes+10}) orbiting the Galaxy since the remote epoch
of its formation. Hence, signatures of such a long-term interaction
could be present in their observational properties.  In this context,
both the inner and the outer portions of the kinematic profiles
provide crucial information on the dynamics of the systems: for
instance, a central gradient in the velocity dispersion profile could
be used to constrain the presence of an intermediate-mass black hole;
the external portion provides information on the possible tidal
perturbations due to interaction of the cluster with the Galactic
tidal field.

A growing set of observational evidence is indeed unveiling an
unexpected dynamical complexity in Galactic GCs, demonstrating that
the traditional assumptions of sphericity, pressure isotropy and
non-rotation are far too simplistic (for references on morphological
distortion, velocity anisotropy and rotation observed in Galactic GCs,
see the comprehensive list reported in \citealp{ferraro+18a}).  In
particular, \citet{fabricius+14} detected signals of rotation in all
the 11 Milky Way GCs studied in their survey and \citet{kamann+18}
found evidence of rotation in $60\%$ of their sample (of 25 objects).
In the context of the MIKiS survey, \citet{ferraro+18a} recently found
rotation signals at distances of a few half-mass radii from the center
in 10 Milky Way GCs (out of 11 investigated) and \citet{lanzoni+18}
detected in M5 one of the cleanest and most coherent rotation patterns
ever observed in a GC.  All these results suggest that, when properly
studied, the vast majority of GCs shows signatures of the presence of
internal rotation. According to a number of numerical studies (see
e.g. \citealt{einsel+99,ernst+07,tiongco+17}), the present-day
signatures could be the relic of a stronger internal rotation set at
the epoch of the cluster's formation (see e.g., \citealp{vesperini+14,
  lee+16, mapelli17}) and gradually altered and erased as result of
the effects of angular momentum transfer and loss due to internal
dynamical processes and star escape. In addition, as shown in the
recent study by \citet{tiongco+18}, the interplay between internal
dynamics and the interaction with the Galactic tidal field can produce
a number of complex kinematical features in rotating clusters,
including a radial variation in the orientation of the rotation axis:
this would be the manifestation of the transition between the inner
regions dominated by the cluster's intrinsic rotation, and the outer
regions dominated by the rotation induced by the Galaxy tidal torque.

As part of the MIKiS survey, here we present the velocity dispersion
and rotation profiles in the intermediate/outer regions of NGC 5986.
This is a relatively poorly investigated Galactic GC (see
\citealt{alves+01} and reference therein). It is massive (with total
$V$-band magnitude $M_V=-8.44$) and of moderate concentration
($c=1.23$), with core and projected half-light (or projected
half-mass, at a first approximation) radii $r_c=28.2\arcsec$ and
$R_h=58.8\arcsec$, respectively (\citealt{harris96}) and with
three-dimensional half-mass radius $r_h\sim 78.16\arcsec$, as obtained
from the corresponding \citet{king66} model.  Because of various
physical properties in common with the group of ``iron-complex'' GCs
\citep{dacosta16}, NGC 5986 was suspected to be the remnant of a
disrupted dwarf galaxy. However, a recent high-resolution
spectroscopic investigation of 25 giant stars \citep{johnson+17}
provided accurate measure of the cluster metallicity ([Fe/H]$=-1.54$)
and showed that it is homogeneous in iron and with a well defined
anti-correlation between the Al and the Mg abundances, in agreement
with what expected for genuine GCs.

The paper is structured as follows, In Section \ref{sec_obs} we
briefly summarize the observations, the adopted data reduction
procedures, and the method used to measure the stellar RVs.  The main
results of our kinematic study are presented in Section
\ref{sec_resu}: the cluster has been characterized in terms of its
systemic velocity and the radial profiles of its ordered and
disordered motions. Section \ref{sec_discuss} discusses the
observational results in the context of recent N-body simulations of
tidally perturbed clusters, illustrating three possible paths that
could lead to the observed features.

\section{Observations, data reduction and radial velocities}
\label{sec_obs}
The observational strategy of the MIKiS survey and the adopted
procedures to reduce the data are presented and discussed by
\citet[][see also \citealt{lanzoni+18}]{ferraro+18a,ferraro+18b}.  In
short, in the case of NGC 5986, we combined data acquired with FLAMES
in the GIRAFFE/MEDUSA mode (using the HR21 grating, with resolving
power R$\sim 16200$ and spectral coverage 8484-9001 \AA) and KMOS
(with the YJ grating, covering 1.00-1.35 $\mu$m at a resolution
R$\approx$3400).  In particular, we secured five pointings with FLAMES
(with integration times ranging from 2400 to 3500 s) and nine KMOS
pointings (with integration times between 30 and 60 s).  The
spectroscopic targets have been selected from the HST-ACS catalog of
\citet{sarajedini+07}, the 2MASS infrared catalog \citep{2MASS} and
ground-based wide-field observations.
Particular care was devoted to select isolated targets with no bright
contaminant neighbor within $2\arcsec$-$3\arcsec$.  The usual
pre-reduction process of the raw data was performed by using the
standard
pipelines.\footnote{http://www.eso.org/sci/software/pipelines/} In the
case of KMOS, the RV of each target was measured from the spectrum
manually extracted from the brightest spaxel of each stellar centroid.

\begin{figure*}
\includegraphics[width=15cm]{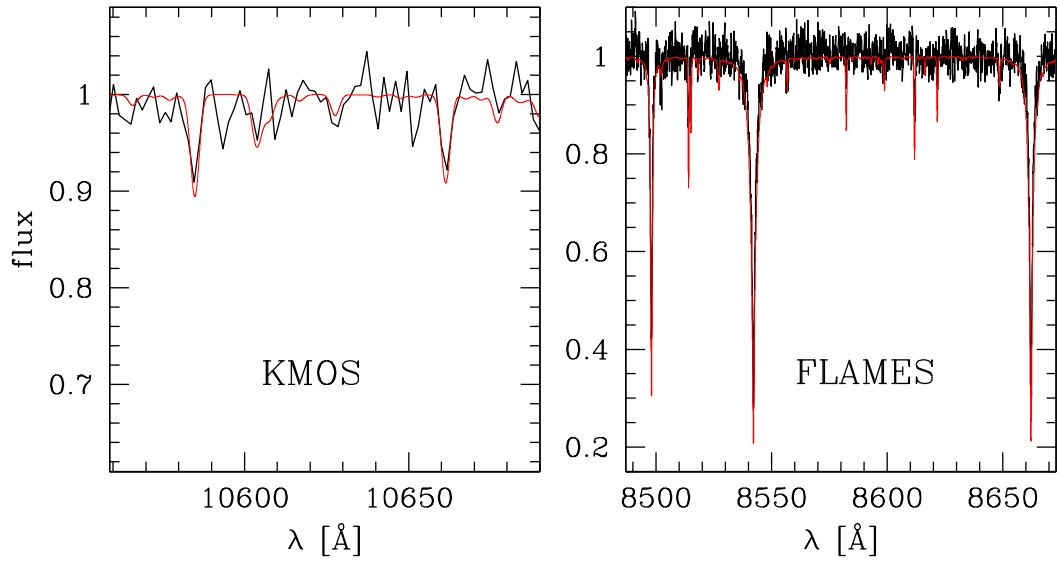}
\caption{Examples of KMOS (left) and FLAMES (right) spectra. The
  observatons are plotted in black, while the synthetic templates are
  in red. Fluxes are arbitrarily normalized to unit.}
\label{fig_spec}
\end{figure*}

RV measurements were obtained following the procedure already
described in previous papers (see \citealp{ferraro+18a,lanzoni+18}).
First, the observed spectra have been corrected for the heliocentric
velocity, then cross-correlated with synthetic spectra of appropriate
metallicity, spectral resolution and stellar parameters.  An example
of synthetic template, and RV-corrected observed spectrum for a KMOS
and for a FLAMES target is shown in Figure \ref{fig_spec}. In the case
of the FLAMES data, the cross-correlation procedure has been applied
to patterns of absorption lines observed in three different wavelength
ranges \citep[see][]{ferraro+18a}. For KMOS data, because of the lower
spectral resolution, and the fact that fewer stellar absorptions and
several telluric lines are present in the near-infrared spectral
region, RVs have been determined from the cross-correlation with
individual features.

For the spectra acquired with FLAMES, typical uncertainties in the RVs
are of the order of 0.5 km s$^{-1}$ and 1 km s$^{-1}$ for the
brightest ($V<16.5$) and the faintest portion of the sample.  These
have been derived from the dispersion of the RV measures obtained from
the considered spectral windows, and also comprise the uncertainty on
the wavelength calibration.  For KMOS, we first verified that no
systematic RV offsets were present with respect to the FLAMES
measures, by using a dozen of stars in common. Then, the RV errors
have been estimated on the basis of the signal-to-noise ratios using
the relation reported in \citet{lanzoni+18}.\footnote{Note that in the
  case of NGC 5986, because of the low metallicity of the cluster,
  several spectra acquired with KMOS had signal-to-noise ratios
  smaller than 30.  Hence, only stars with errors below to 9 km
  s$^{-1}$, with a mean value of 6.8 km s$^{-1}$, have been retained
  and used in the present analysis.} For the stars measured with both
the instruments, the RV determined from the FLAMES spectra has been
adopted.  The kinematic catalog of NGC 5986 is freely available at the
MIKiS web page.\footnote{
  http://www.cosmic-lab.eu/Cosmic-Lab/MIKiS\_Survey.html}

\section{Results}
\label{sec_resu}
\subsection{Systemic velocity}
\label{sec_vsys}
\begin{figure}
\includegraphics[width=15cm]{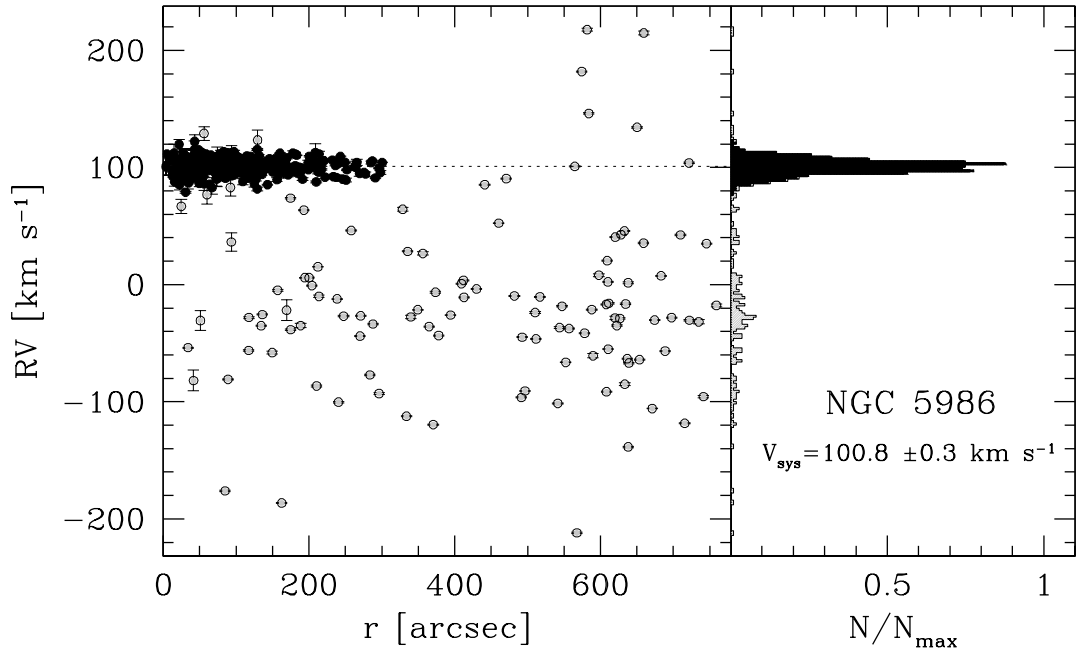}
\caption{{\it Left panel:} Radial distribution of the line-of-sight
  velocities measured from the MIKiS observations in NGC 5986.  The
  selected cluster members are plotted as black circles, while field
  contaminants are in grey. {\it Right panel:} RV distribution as a
  normalized histogram. $V_{\rm sys}$ denotes the cluster systemic
  velocity.}
\label{fig_vrr}
\end{figure}

\begin{figure}
\includegraphics[width=15cm]{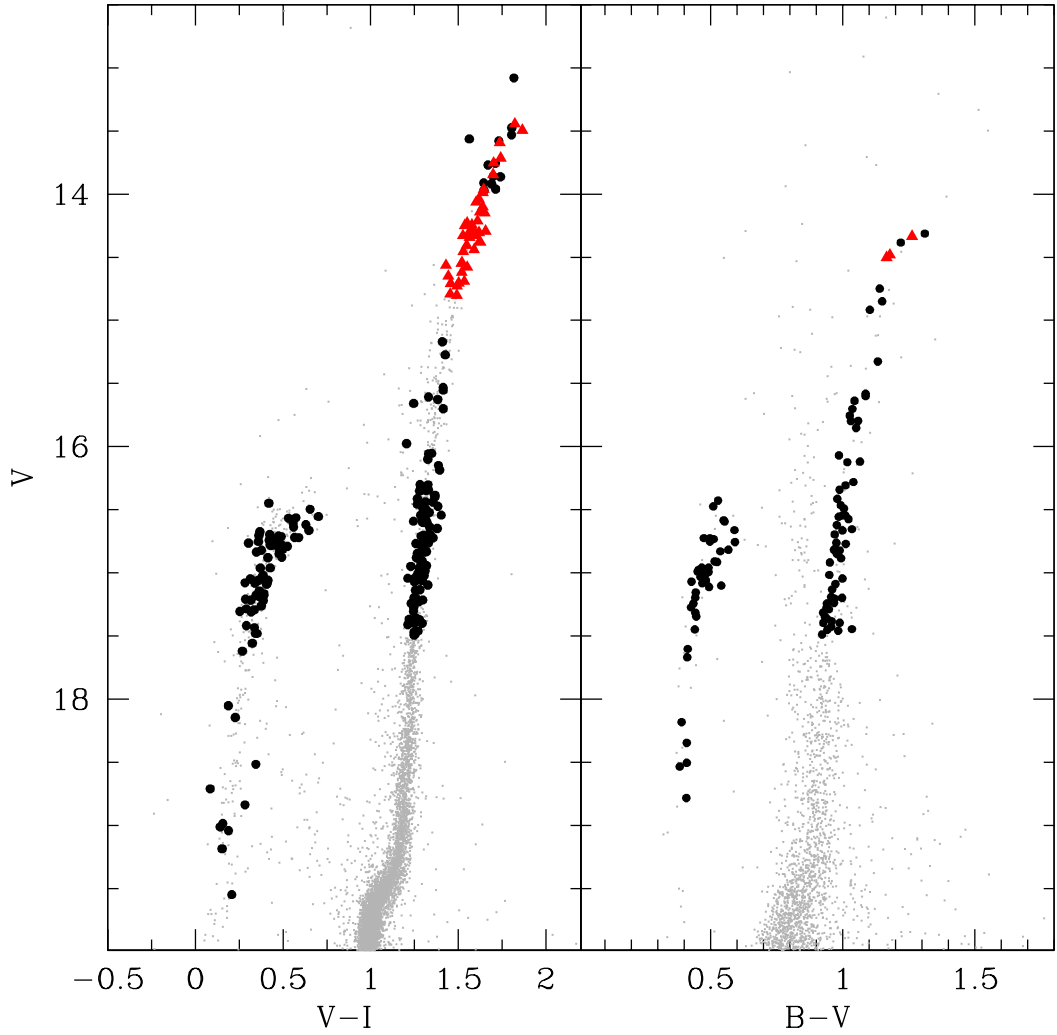}
\caption{Position of the surveyed kinematic members in the
  color-magnitude diagrams of NGC 5986. The KMOS targets are
  highlighted as red triangles, while the stars surveyed with FLAMES
  are plotted as black circles.}
\label{fig_cmd}
\end{figure}

\begin{figure}
\includegraphics[width=15cm]{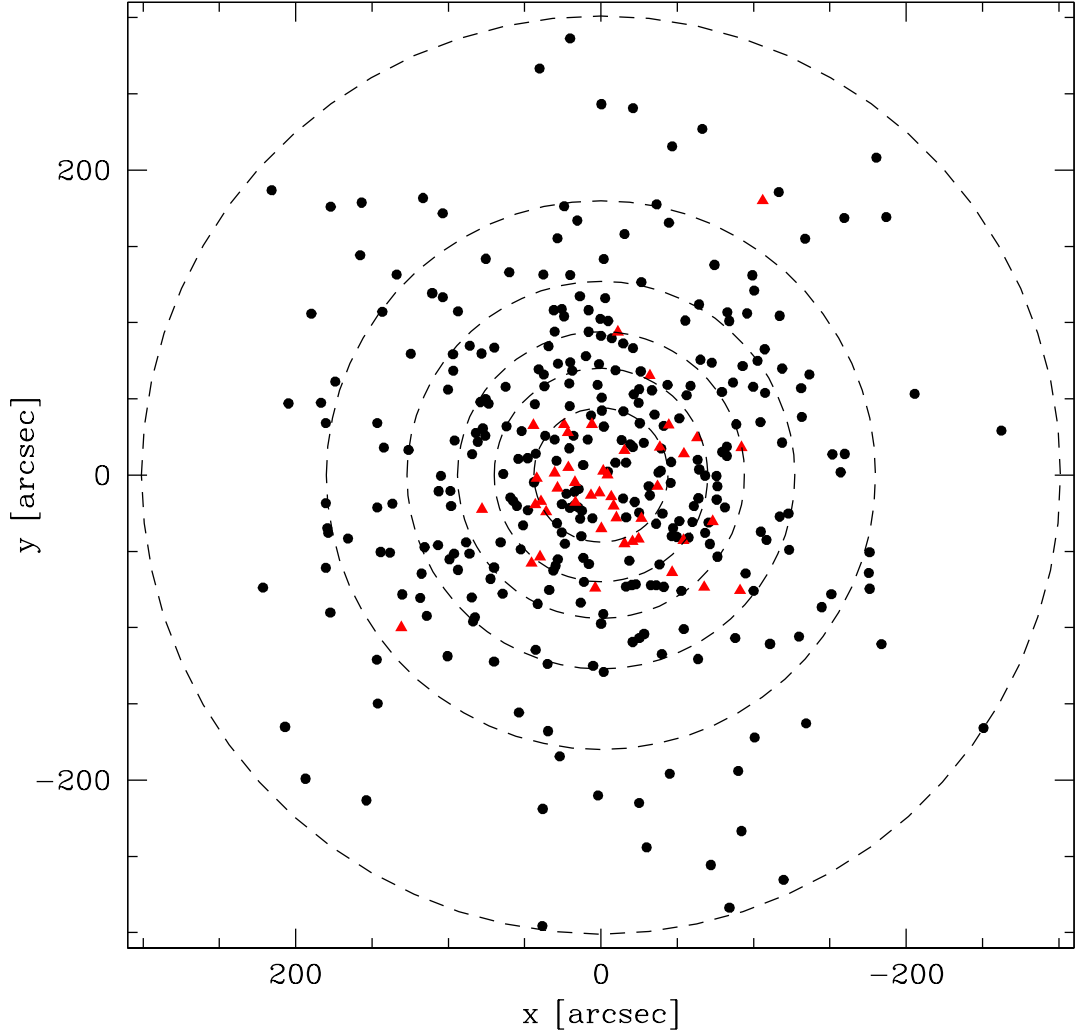}
\caption{Position the surveyed kinematic members in the plane of the
  sky. Symbols have the same meaning as in Figure \ref{fig_cmd}. The
  plotted coordinates, x and y, are the right ascension and
  declination relative to the cluster center, determined by using
  equation (1) in \cite{vandeven+06}. The dashed circles mark the
  boundaries of the radial bins used to determine the cluster velocity
  dispersion (see Table \ref{tab_vd}).}
\label{fig_map}
\end{figure}

We illustrate in Figure \ref{fig_vrr} the radial distribution of the
470 line-of-sight velocities measured in this work.  As value for the
cluster center we adopted the one quoted in \citet{goldsbury+11}.
Although the surveyed stars are located out to $r=760\arcsec$, cluster
members appear to be distributed within $\sim 300\arcsec$ from the
center and are all grouped around a mean velocity of $\sim 100$ km
s$^{-1}$.  The remaining objects clearly are Galactic field
contaminants, showing a much broader RV distribution. To determine the
cluster systemic velocity ($V_{\rm sys}$) we conservatively used only
the sub-sample of (313) likely-member stars observed with FLAMES at
$r<300\arcsec$.  After applying 3-$\sigma$ rejection cleaning to this
sample, we estimated $V_{\rm sys}$ by following the maximum-likelihood
approach fully described in Section 3.4.1 of \citet[][see also
  \citealp{martin+07, sollima+09}]{walker06}. Briefly, this consists
in determining the values of the mean velocity and velocity dispersion
that maximize the joint probability of having observed a given set of
RVs, under the hypothesis that the measures have a Gaussian
distribution (which is fully appropriate for this cluster; see Figure
\ref{fig_vrr}). By construction, this method also takes into account
the error associated to each individual RV. To estimate the
uncertainties on the two parameters, we have followed the classical
approach outlined by \citet{pryor+93}. The mean velocity thus
obtained, corresponding to the cluster systemic velocity, is $V_{\rm
  sys}=100.8 \pm 0.3$ km s$^{-1}$.  This is larger than the value
quoted by \citet[][2010 edition: $V_{\rm sys}=88.9\pm 3.7$ km
  s$^{-1}$]{harris96}, but in very good agreement with the recent
estimate of \citet[][$V_{\rm sys}=99.76$ km s$^{-1}$]{johnson+17}.  In
the following, we will denote as $V_r\equiv {\rm RV} -V_{\rm sys}$ the
RVs to which the cluster systemic velocity has been subtracted.
Figures \ref{fig_cmd} and \ref{fig_map} show the color-magnitude
diagram and the distribution on the plane of the sky of the selected
cluster members. As apparent, the KMOS targets are all bright giants
($V<14.8$) mainly located in the cluster central region, while the
stars surveyed with FLAMES are observed along both the red giant
branch down to $V=17.5$ and the horizontal branch, sampling both the
center and the outskirts of the system.

\subsection{Rotation curve and velocity dispersion profile}
\label{sec_kin}
We searched for signatures of ordered motions in NGC 5986 following
the same approach adopted in the previous papers of this series
\citep{ferraro+18a, lanzoni+18} and presented in full detail by, e.g.,
\citet[][see also \citealp{lanzoni+13}]{bellazzini+12}. To define the
position angle (PA) of the rotation axis we considered a line passing
through the center of the system and dividing the observed RV sample
in two portions. We then varied the PA of the line in the
anti-clockwise direction from $0\arcdeg$ (North) to $180\arcdeg$
(South), at each step determining the difference $\Delta V_{\rm mean}$
between the average velocity of the RV samples on either sides of the
line. Finally, we fitted the variation of $\Delta V_{\rm mean}$ as a
function of PA with a sine function.  In the presence of ordered
motions, the rotation amplitude ($A_{\rm rot}$) is equal to half the
maximum absolute value of the sine function, while the rotation axis
position angle (PA$_0$) corresponds to the abscissa of the maximum.
Moreover, the cumulative RV distributions of the samples located on
either sides of the best-fit rotation axis, as well as the mean RV of
the two sub-groups are expected to be different.  Three indicators
have thus been used to verify whether these differences are
statistically significant: we evaluated the probability that the two
RV distributions are extracted from the same parent family through a
Kolmogorov-Smirnov test; we tested the hypothesis of different sample
means through the Student's t-test and a maximum-likelihood approach.

\begin{figure}
\includegraphics[width=15cm]{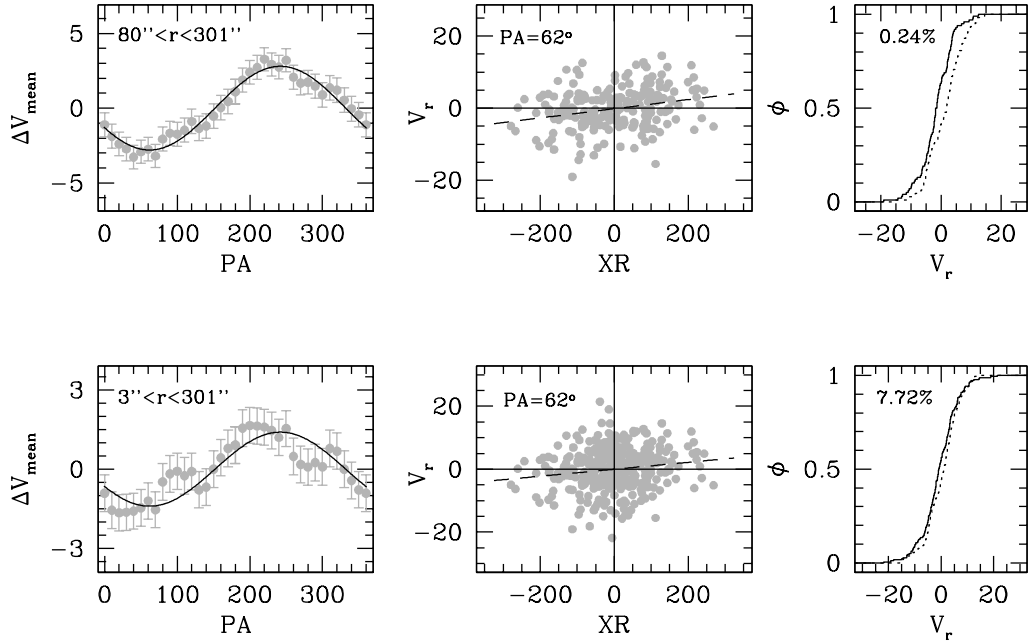}
\caption{Diagnostic diagrams (see Section \ref{sec_kin}) of the
  rotation signatures detected at $r>80\arcsec$ (upper row) and over
  the entire radial range surveyed (lower row).  {\it Left panels:}
  difference between the mean line-of-sight velocity calculated in the
  two half-planes, as a function of the position angle of the dividing
  line.  The labels indicate the radial range considered.  {\it
    Central panels:} spatial distribution of the radial velocities (in
  km s$^{-1}$). XR denotes the projected distance from the best-fit
  rotation axis (in arcsec), with the value of the position angle
  marked.  {\it Right panels:} cumulative distribution of the RVs with
  positive (dotted line) and negative (solid line) projected distance
  XR. The outcome of a KS test of the two distributions is reported in
  the top left corner of each panel.}
\label{fig_vrot}
\end{figure}

By applying this procedure in distinct circular annuli around the
cluster center, we found a significant signature at $r>80\arcsec$ (see
the top panels of Figure \ref{fig_vrot} and the values listed in the
first row of Table \ref{tab_vrot}).  Indeed, all the three adopted
estimators indicate that the detected rotation signal has high
statistical significance: a Kolmogorov-Smirnov test returns a
probability $P_{\rm KS}=2.5 \times 10^{-3}$ that the two sub-samples
are originated from the same parent distribution, following the
Student's t-test, the probability that the means of the two samples
are different is larger than 99.8\% and, based on the
maximum-likelihood approach, this difference is significant at 3.4
$\sigma$.  If we apply the same procedure to the whole sample of stars
(from $3\arcsec$ to $300\arcsec$) fixing the rotation axis at the
best-fit position angle obtained for $r>80\arcsec$ (PA$_0=62\arcdeg$),
we find the results plotted in the bottom panels of Figure
\ref{fig_vrot} and listed in the second row of Table \ref{tab_vrot}.
The significance of the global signal is weaker than that obtained for
$r>80\arcsec$ since the data considered for this calculation include
the cluster's innermost regions where the strength of the rotation
becomes increasingly smaller.

\begin{figure}
\includegraphics[width=15cm]{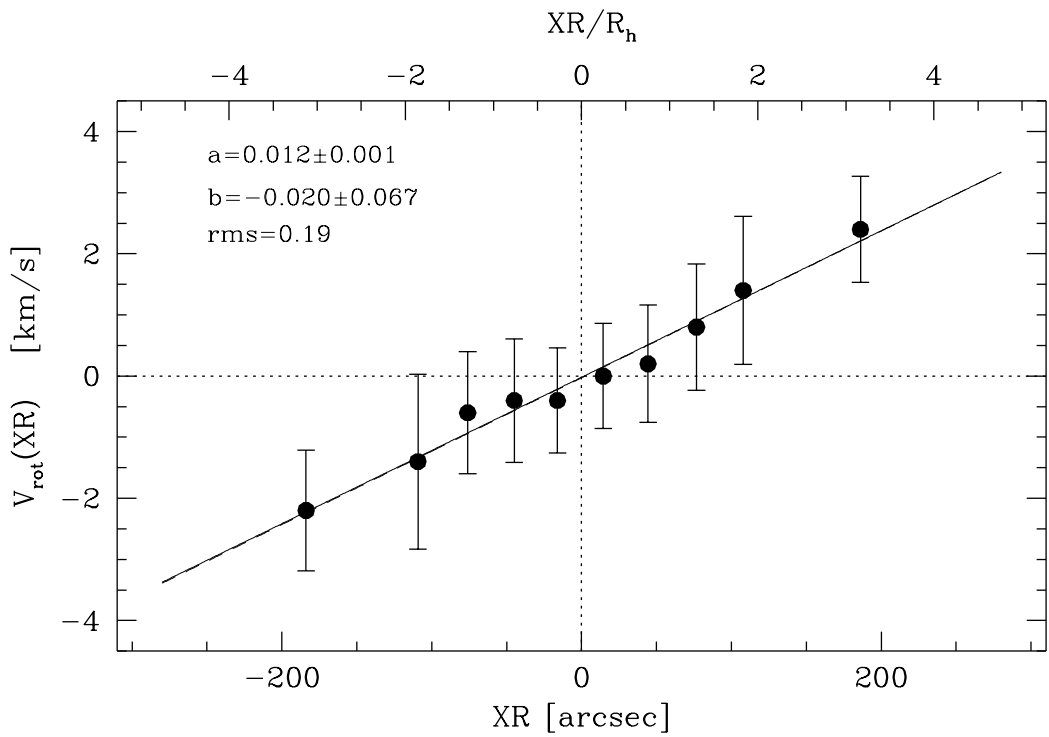}
\caption{Rotation curve of NGC 5986.  The mean velocity as a function
  of the distance from the rotation axis is displayed (see also Table
  \ref{tab_kin}). The best-fit line to the observed behavior has slope
  ($a$), intercept ($b$) and rms scatter (rms) as labelled in the
  figure. }
\label{fig_rotcurve}
\end{figure}

To determine the rotation curve of NGC 5986 we have first rotated the
Cartesian coordinate system (x,y) over the position angle PA$_0$.  By
construction, the resulting coordinate system (XR,YR) has the XR axis
along the cluster major axis, and YR aligned with the rotation
axis. We then applied the same maximum-likelihood method as above
(Section \ref{sec_vsys}) to determine the stellar mean velocity in
five intervals of XR on both sides of the rotation axis.  The rotation
curve obtained through this procedure is plotted in Figure
\ref{fig_rotcurve} and listed in Table \ref{tab_kin}.  It shows a
clear solid-body rotation pattern, monotonically increasing from 0 km
s$^{-1}$ in the center, up to $\sim 2$ km s$^{-1}$ at the outermost
sample distances, roughly corresponding to 4 half-mass radii.  Its
least square best-fit line has equation $V_{\rm rot} = a \times {\rm
  XR} +b$, where $a=0.012$ km s$^{-1}$ arcsec$^{-1}$ and $b=-0.02$ km
s$^{-1}$, and high statistical significance, with a the Spearman rank
correlation coefficient $\rho = 0.99$.

In principle, a non-negligible amount of the detected rotation could
be due to the line-of-sight component of the cluster space motion
\citep[see, e.g.,][]{feast+61}.  We thus checked the amplitude of the
perspective rotation by following \citet{vandeven+06}. As systemic
velocity we adopted the one determined in Section \ref{sec_vsys}. For
the systemic proper motion we used both the value quoted in
\citet{casetti+07} and the one recently measured from the second Gaia
data release \citep{helmi+18}.  In both cases, we found that the
cluster perspective rotation is negligible.

\begin{figure}
\includegraphics[width=15cm]{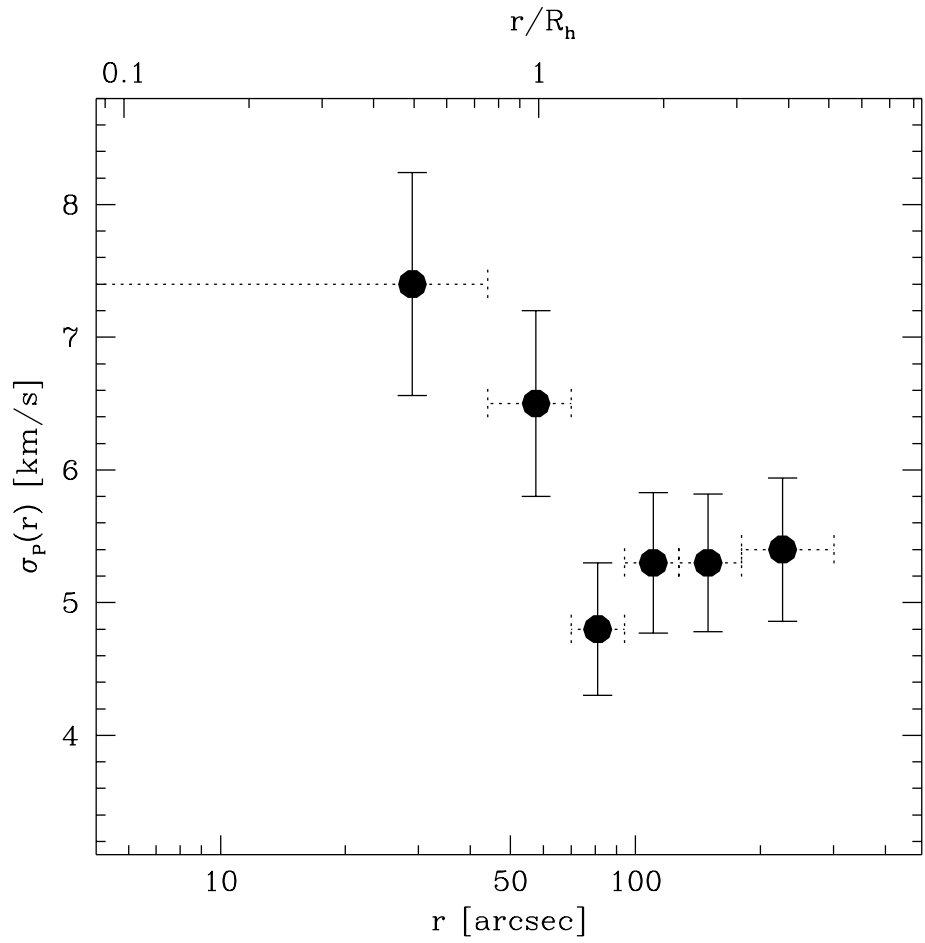}
\caption{Projected velocity dispersion profile, determined after
  subtraction of the rotational contribution to the individual RVs, in
  six concentric annuli around the cluster center.}  
\label{fig_vd}
\end{figure}

To determine the projected velocity dispersion profile $\sigma_p(r)$
of the cluster, we have first subtracted from each individual RV the
mean rotational velocity at the position of the star. This removes the
contribution of rotation to the second velocity moment, thus providing
the ``pure'' velocity dispersion measure.  Then, we applied the
maximum-likelihood approach discussed above (Section \ref{sec_vsys})
in six circular bins around the center of the system, and determined
$\sigma_p(r)$ as the velocity dispersion value that maximizes the
joint probability of having observed the set of RVs (and their
associated errors) measured in that radial range.  The resulting
velocity dispersion profile is listed in Table \ref{tab_vd} and
plotted in Figure \ref{fig_vd}.  As detailed in the table, the FLAMES
measures (having errors of $\sim 1$ km s$^{-1}$) always dominate the
samples, both in the most central bin and, especially, in the
outermost radial annuli (indeed, the obtained profile is perfectly
consistent within the errors with that derived by using only the
FLAMES measures). As apparent from the figure, after a decline between
$0\arcsec$ and $\sim 90\arcsec$, the velocity dispersion stays
approximately constant at $\sim 5$ km s$^{-1}$ out to the outermost
sampled radius.

\section{Summary and Discussion}
\label{sec_discuss}
In this paper we presented a novel investigation of the internal
kinematics of the Galactic GC NGC 5986, as based on hundred spectra of
individual stars.  We find one of the most significant rotation
patterns hitherto detected in GCs (see also the cases of NGC 4372 in
\citealt{kacharov+14}, 47 Tucanae in \citealt{bellini+17}, and M5 in
\citealp{lanzoni+18}), the only one so far showing a clear solid-body
rotation behavior. For the first time, we also jointly find evidence
of a velocity dispersion profile that flattens at distances larger
than about one half-mass radius. The physical interpretation of the
co-existence of these kinematic features requires particular care,
because several dynamical pathways may be envisaged as a possible
origin of the observed behaviors. A qualitative discussion of three
possible channels, as based on recent theoretical investigations into
the kinematic evolution of collisional stellar systems, is provided
below.

\begin{figure}
\includegraphics[angle=0,width=13cm]{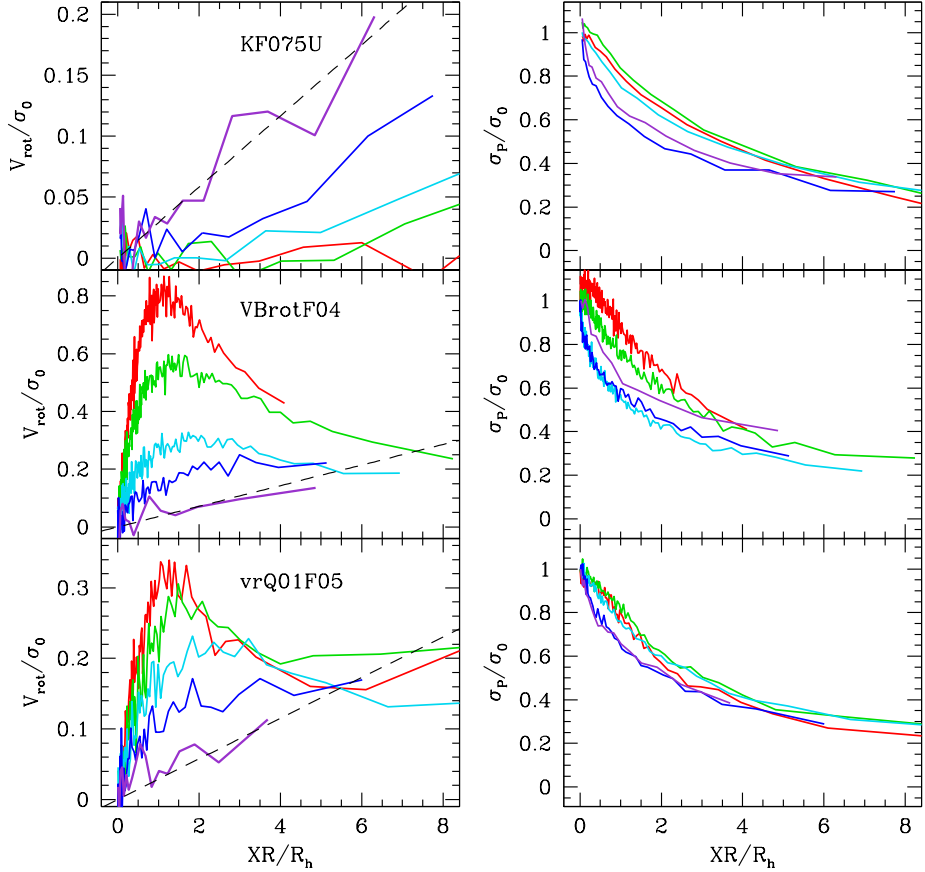}
\caption{Time evolution of the rotation curve (left-hand panels) and
  velocity dispersion profile (right-hand panels), both calculated
  from line-of-sight velocities and normalized to the central velocity
  dispersion, for three of the N-body simulations from
  \citet{tiongco+16a}. The radius is normalized to the projected
  half-mass radius, $R_{\rm h}$. Different lines correspond to
  different dynamical stages, with the red, green, cyan, blue and
  purple lines corresponding to increasing dynamical ages. In the left
  panels, a dashed line shows a theoretical solid-body rotation curve
  with slope $\Omega/2$, a value that star clusters will converge to
  during their long-term evolution in a tidal field (see
  \citealp{tiongco+16b}). In all cases, both sets of kinematic
  profiles have been calculated by considering all stars enclosed
  within one Jacobi radius.}
\label{fig_models}
\end{figure}

One possibility is that such kinematic properties result from the
evolution of the cluster in an external tidal field, with attention to
the role played by a population of energetically unbound stars
confined within the critical equipotential surface (``potential
escapers''; see \citealt{heggie01}). Recent studies based on direct
N-body simulations have shown that such a population determines a
flattening of the velocity dispersion profile in the outer regions of
a stellar system \citep{kuepper+10,claydon+17}. In addition, tidally
perturbed clusters, during the course of their dynamical evolution
(and irrespectively of their initial conditions), tend to develop a
signature of solid-body rotation which depends on their orbital
angular velocity partial synchronization (see
\citealt{tiongco+16b,claydon+17}) and a degree of anisotropy which
depends on their initial filling factor (e.g., see \citealt{giersz+97,
  tiongco+16a}).  As a representative example of such an evolutionary
behaviour, we illustrate the rotation curve and velocity dispersion
profile of an N-body model originally presented by
\citet{tiongco+16b}, depicted a different times (see Figure
\ref{fig_models}, first row). Such an N-body simulation starts from
initial conditions sampled from a \citet{king66} equilibrium model,
which is initially non-rotating, and is evolved on a circular orbit in
the tidal field of a Keplerian potential (with an initial ratio of the
intrinsic half-mass radius to the Jacobi radius given by
$r_h/r_J=0.087$). As apparent from the figure, the system
progressively develops a solid-body rotation curve with an angular
velocity consistent with $\Omega/2$ (where $\Omega$ denotes the
orbital angular velocity; for further details, see model ‘KF075U’ in
\citealt{tiongco+16b}). The corresponding velocity dispersion profiles
progressively flatten in the outermost regions of the system (see
right-hand panels).

A second pathway corresponds to the case of a collisional system which
is initially characterized by some intrinsic internal rotation and is
progressively evolving towards a condition of solid-body rotation
(starting from its central to intermediate regions), as a result of
the angular momentum transport and loss, induced by two-body
relaxation processes. Such a case is exemplified by an N-body model
originally presented by \citet{tiongco+16a}, for which we again
illustrate the evolution of the rotation curve and velocity dispersion
profile at different times (see Figure \ref{fig_models}, second row).
The simulation starts from initial conditions sampled from
differentially rotating equilibrium models \citep{varri+12} and
includes the effect of a mild tidal field (corresponding to an initial
ratio of the intrinsic half-mass radius to the Jacobi radius
$r_h/r_J=0.093$) associated to a circular orbit in a Keplerian
galactic potential (for further details, see model VBrotF04 in
\citealt{tiongco+16a}). In this case, the evolution of the central
slope value of the rotation curve is not determined by the influence
of the tidal environment. In the represented model, only at a very
late stage of evolution the rotation curve settles, as in the previous
case, into a solid body-like behavior, with an angular velocity
determined by the condition of partial synchronization.  We wish to
note that, in this pathway, the orientation of the axis of the initial
intrinsic rotation, in principle, may also be different from the
orientation of the orbital rotation axis \citep[see][]{tiongco+18}. As
a result, a radial variation of the kinematic position angle may be
found \citep[e.g., see][]{bianchini+13,boberg+17}, which is not
observed in the range explored in the current kinematic study.

An additional possible dynamical path is that of a stellar system
experiencing a phase of ``violent relaxation'' in the presence of an
external tidal field; in this case the violent relaxation process and
the effects of the tidal field leave a distinctive fingerprint on the
cluster's internal kinematics from the very early stages of its
evolution (see \citealt{vesperini+14}), with a solid-body like
behavior of the central portion of the rotation curve. Also in this
case, the differential rotation imprinted during the early
evolutionary stages progressively evolves towards the condition of
partial synchronization described in the two previous cases. This
third case is exemplified by a model originally presented by
\citet{tiongco+16a} and subsequently studied by \citet[][see model
  vrQ01F05 for further details]{tiongco+17}. The element of interest
in this third case (see Figure \ref{fig_models}, third row) mostly
concerns the velocity dispersion profile, which becomes more flattened
and extended at earlier times compared to the previous two cases. Such
a difference is due to the fact that this model develops the
population of energetically unbound stars responsible for the
flattening of the velocity dispersion during the early violent
relaxation phase.

The models shown in Figure \ref{fig_models} serve as a framework for a
general interpretation of the observed features, but are not designed
to provide a detailed and quantitative fit to the kinematical
properties of NGC 5986. In particular we note that the flattening in
the velocity dispersion observed in NGC 5986 is more extreme and
internal than that found in all our models. As shown in several
studies \cite[see][]{casetti+07, allen+08, moreno+14}, NGC 5986 is on
an eccentric orbit and specific models for this cluster need to take
into account the effects on the velocity dispersion due to the time
variation in the strength of the external tidal field \citep[see,
  e.g.,][]{kuepper+10}. Indeed, a time-variable tidal field, as
experienced by a star cluster moving on an elliptic orbit, has direct
implications on the properties and time evolution of the population of
potential escapers dynamically generated in a collisional stellar
system, which, in turn, has an impact on the shape of its outer
velocity dispersion profile (see especially Figure 3 in
\citealp{kuepper+10}, and also \citealp{drukier+07, claydon+17}).  We
estimated the orbital parameters of the system in the
\citet{johnston+95} Milky Way potential well, adopting $V_{\rm sys} =
100.8$ km s$^{-1}$ as systemic line-of-sight velocity (see Section
3.1) and the proper motions of \citet{casetti+07} for the two
components on the plane of the sky. We found that the system has a
highly eccentric orbit ($e = 0.80$), practically plunging into the
central part of the Milky Way along a path that is confined within the
innermost few kpc from the Galactic center, with a pericentric
distance $r_p = 0.5$ kpc and the apocenter at $r_a = 5.4$ kpc. These
values are in very good agreement with those determined by
\citet{casetti+07}, who used only a slightly smaller value of the
systemic velocity. The orbital parameters estimated by
\citet{helmi+18} from the proper motions presented in the second Gaia
data release also suggest that this cluster is on a very eccentric
orbit with a very small pericentric distance: depending on the model
adopted for the Galactic potential, the eccentricity is equal to 0.7,
0.81 or 0.87, while $r_p$ varies between 0.85, 0.52 and 0.07 kpc,
respectively.  The orbital parameters derived for NGC 5986 thus
suggest quite intense interactions with the central regions of the
Galaxy, with an orbital radial period of only $\sim 60$ Myr,
corresponding to a few hundreds passages of the cluster close to the
Galactic center during its lifetime ($t=12$ Gyr; \citealt{forbes+10}).
Thus, the effect of the time variation in the external tidal field is
certainly playing an important role in the dynamics of NGC 5986 and
must be included in simulations aimed at providing a detailed fit of
the observed properties of this cluster. We point out here that, if
the solid-body rotation revealed by our observations is the result of
the cluster's convergence toward a state of partial synchronization
with angular velocity $\Omega/2$, the angular velocity measured for
NGC 5986 would correspond to the value of $\Omega/2$ at about 0.46 kpc
(simply calculated assuming a circular velocity equal to 220 km/s).

To provide a comprehensive dynamical interpretation of this cluster
some additional pieces of information are needed.  First of all, the
line-of-sight kinematics should be assessed also in the outer regions
of the cluster, in the proximity (and ideally beyond) the nominal
Jacobi radius. Indeed, the shape of the outer rotation curve is
crucial to test the applicability of the three pathways described
above.  We also plan to search for density distortion or streams
associated with tidal perturbations.  In principle, crucial
information will also be added by the stellar proper motions measured
by Gaia that, once combined with the line-of-sight kinematics, would
allow us to reconstruct the full three-dimensional structure of the
velocity space of the system.  We have already started to analyze the
most recent Gaia data \citep[see, e.g.,][]{helmi+18}, but this study
is particularly challenging for NGC 5986 because of its relatively
large distance from Earth (10.4 kpc; \citealp{harris96}), the high
level of Galactic contamination in the cluster's direction on the sky,
and the fact that the cluster proper motion is almost
indistinguishable from that of the bulge.  From the theoretical point
of view, we plan to construct dynamical models and N-body simulations
specifically tailored to the case of NGC 5986, exploring, in
particular, the effects of an external tidal field in the case of
highly eccentric orbits repeatedly crossing the innermost region of
the Galaxy.  Such a comprehensive investigation is particularly
timely, in light of the growing interest for the physical
understanding of the morphological and dynamical properties of the
peripheries of Galactic GCs, especially regarding the interpretation
of the kinematic properties of possible ``extra-tidal'' structures.

\acknowledgements{FRF acknowledges the ESO Visitor Programme for the
  support and the warm hospitality at the ESO Headquarter in Garching
  (Germany) during the period when part of this work was performed.}

\newpage
\begin{table*}[h!]
\caption{Rotation Signatures in NGC 5986}
\begin{tabular}{rrrrccccc}
  & & & & & & & & \\
\hline
$r_i$  & $r_e$ & $r_m$ &  $N$ & PA$_0$ & $A_{\rm rot}$ & $P_{\rm KS}$ & $P_{\rm Stud}$ & n-$\sigma_{\rm ML}$\\
\hline
   80  &  301  & 149.5 &  209 &   62  & 1.4 & $2.5 \times 10^{-3}$ &  $>99.8$  & 3.4 \\ 
   3  &  301  & 107.7 &  358 &   62  & 0.7 & $7.7 \times 10^{-2}$ &  $>90.0$  & 2.2 \\
\hline
\tablecomments{Rotation signatures detected at $r>80\arcsec$ (first
  row) and over the entire radial range surveyed by the observations
  (second row). Starting from the left: inner ($r_i$) and outer
  ($r_e$) radius of the considered radial interval (arcseconds), mean
  distance from the center of the stars in the annulus ($r_m$), star
  number ($N$), position angle of the rotation axis (PA$_0$, in
  degrees), rotation amplitude of the best fitting sinusoidal function
  (A$_{\rm rot}$, in km s$^{-1}$; see Fig. \ref{fig_vrot}),
  probability outcome of the KS test ($P_{\rm KS}$) and of the
  t-Student test ($P_{\rm Stud}$), significance (n-$\sigma_{\rm ML}$)
  of the difference of the two means according to a maximum-likelihood
  approach (units of n-$\sigma$).}
\end{tabular}
\label{tab_vrot}
\end{table*}

~

~

~

\begin{table}[h!]
\begin{center}
\caption{Rotation Curve of NGC 5986}
\begin{tabular}{rrrrccrrcc}
 & & & & & & & & & \\
\hline
XR$_i$ & XR$_e$ & XR$_m+$ &  $N+$ & $V_{\rm rot}+$ & $\epsilon_{V+}$ & XR$_m-$ &  $N-$ & $V_{\rm rot}-$ & $\epsilon_{V-}$ \\
\hline
   1 &  30 &  14.5 &  51 &  0.0 & 0.9 &  -16.1 &  56 & -0.4 & 0.9 \\ 
  30 &  60 &  44.4 &  45 &  0.2 & 1.0 &  -44.9 &  41 & -0.4 & 1.0 \\
  60 &  90 &  76.7 &  32 &  0.8 & 1.0 &  -76.0 &  26 & -0.6 & 1.0 \\
  90 & 130 & 107.9 &  27 &  1.4 & 1.2 & -109.2 &  25 & -1.4 & 1.4 \\
 130 & 310 & 186.1 &  25 &  2.4 & 0.9 & -183.8 &  26 & -2.2 & 1.0 \\
\hline
\end{tabular}
\tablecomments{Rotation curve of NGC 5986 in a coordinate system (XR,
  YR) where the vertical axis is oriented as the rotation axis and
  points toward the North-East direction, while XR points toward
  North-West. The first two columns report the inner (XR$_i$) and
  outer (XR$_e$) boundaries (arcseconds) of the five considered bins.
  Columns 3-6 list, for the positive side of the XR axis, the mean
  cluster-centric distance (XR$_m+$) of the stars in each bin, their
  number ($N+$), and their average velocity and associated error in km
  s$^{-1}$ ($V_{\rm rot}+$ and $\epsilon_{V+}+$, respectively).  The
  quantities reported in columns 7-10 are equivalent to those of 3-6,
  for the bins in the negative side of the XR axis.}
\label{tab_kin}
\end{center}
\end{table}

\newpage
\begin{table}[h!]
\begin{center}
\caption{Projected Velocity Dispersion Profile of NGC 5986}
\begin{tabular}{rrrrrrcc}
  & & & & & & & \\
\hline
$r_i$ & $r_e$ & $r_m$ & $N$ & $N_K$ & $N_F$ & $\sigma_P$ &  $\epsilon_{\sigma}$ \\
\hline
   3 &  44 &  29.0 &  61 &  23 & 38 & 7.4 & 0.8 \\
  44 &  70 &  57.6 &  60 &  10 & 50 & 6.5 & 0.7 \\
  70 &  94 &  81.3 &  59 &   7 & 52 & 4.8 & 0.5 \\
  94 & 127 & 110.3 &  61 &   3 & 58 & 5.3 & 0.5 \\
 127 & 180 & 149.5 &  59 &   1 & 58 & 5.3 & 0.5 \\
 180 & 301 & 226.0 &  57 &   1 & 56 & 5.4 & 0.5 \\
\hline
\end{tabular}
\tablecomments{Projected velocity dispersion profile of NGC 5986,
  calculated in circular annuli around the center of the
  cluster. Starting from the left, the columns report: internal
  ($r_i$) and external ($r_e$) boundaries of the annuli (arcseconds),
  average cluster-centric position of the stars belonging to the bin
  ($r_m$), total number of stars ($N$), number of KMOS and FLAMES
  targets ($N_K$ and $N_F$, respectively), projected velocity
  dispersion ($\sigma_P$) and associated error ($\epsilon_\sigma$), in
  km s$^{-1}$.}
\label{tab_vd}
\end{center}
\end{table}

\newpage
 
\end{document}